\documentclass[10pt,conference]{IEEEtran}
\IEEEoverridecommandlockouts\IEEEpubid{\makebox[\columnwidth]{\copyright~2018 IEEE \hfill} \hspace{\columnsep}\makebox[\columnwidth]{ }}

\usepackage[utf8]{inputenc}
\usepackage[T1]{fontenc}
\usepackage{tabularx}
\usepackage{import}
\usepackage{multirow}
\usepackage{todonotes}
\usepackage{longtable}
\usepackage{lscape}
\usepackage{enumitem}
\usepackage{rotating}
\usepackage{microtype}
\usepackage{graphicx}
\usepackage{subfigure}
\usepackage{amssymb}
\usepackage{mathtools}
\usepackage{pifont}
\usepackage[final]{listings} 
\usepackage{color}
\usepackage{algorithm2e}
\usepackage{soulutf8}
\usepackage{epigraph}
\usepackage{tikz}
\usepackage[skip=2pt]{caption}

\usepackage{cite}
\usepackage{amsmath,amssymb,amsfonts}
\usepackage{algorithmic}
\usepackage{graphicx}
\usepackage{textcomp}
\usepackage[english]{babel}

\usepackage{url}

\newcommand{\etal}{\mbox{et~al.}\ }

\usepackage{amssymb}

\usepackage{tikz}
\usetikzlibrary{shapes}
\usetikzlibrary{calc}
\tikzset{myptr/.style={-{Latex[scale=1.5]}}}

\tikzset{MyRoundedBox/.style={
		draw,
		rounded corners=3pt,
		inner sep=5pt,
		align=center
	}
}

\usepackage{listings}
\usepackage{cite}

\usepackage[binary-units=true]{siunitx}

\mathchardef\mhyphen="2D 

\usepackage{booktabs}
\usepackage{array}
\usepackage{multirow}

\usepackage{pifont}
\usepackage{expdlist}

\newcommand{\flex}{FlexSMC\xspace}
\makeatletter
\newcommand{\customlabel}[2]{%
   \protected@write \@auxout {}{\string \newlabel {#1}{{#2}{\thepage}{#2}{#1}{}} }%
}
\makeatother
\newcounter{Req}
\newcommand{\req}[1]{%
\customlabel{#1}{R.\theReq}%
(R.\theReq)%
\stepcounter{Req}%
}

\newcommand{\FRESCO}{\textsc{Fresco}\xspace}

\newcommand{\image}[3]{
\message{Including diagram: #1}
\begin{figure}
\resizebox{\columnwidth}{!}{%
\includegraphics{#1.pdf}
}
\caption{#2}
\label{#3}
\end{figure}
}

\begin{document}
%

\title{A Management Framework for Secure Multiparty Computation in Dynamic Environments
\thanks{
This work has been supported by the German Federal Ministry of Education
and Research, project DecADe, grant 16KIS0538
and the German-French Academy for the Industry of the Future.
}
}

\author{\IEEEauthorblockN{ Marcel von Maltitz, Stefan Smarzly, Holger Kinkelin and Georg Carle}
\IEEEauthorblockA{ Technische Universität München, Department of Informatics \\ Chair for Network Architectures and Services
\\
85748 Garching b. München, Germany \\
\{lastname\}@net.in.tum.de}

}


\maketitle

\newif\iflongversion
\longversiontrue
\longversionfalse

\newenvironment{addendum}{\begin{color}{blue}}{\end{color}}

\newcommand{\branch}[2]{\iflongversion{}#1\else{}#2\fi}
\newcommand{\onlylong}[1]{\branch{#1}{}}

\begin{abstract}
  Secure multiparty computation (SMC) is a promising technology for privacy-preserving collaborative computation\onlylong{ which has been investigated for several decades}.
  In the last years several feasibility studies
  have shown its practical applicability in different fields.\onlylong{ Research is ongoing to improve security and performance properties.}
However, it is recognized that administration, and management overhead of SMC solutions are still a problem \cite{SEPIANetwork} \cite{SMCFinancial}.
  A vital next step is the incorporation of SMC in
  the emerging fields of the Internet of Things and (smart) dynamic environments.
  In these settings, the properties of these contexts make utilization of SMC even more challenging since some vital premises
  for its application regarding environmental stability and preliminary configuration are not initially fulfilled.

  We bridge this gap by providing \flex, a management and orchestration framework for SMC which supports the discovery of nodes, supports a trust establishment between them and realizes robustness of SMC session by handling nodes failures and communication interruptions.
  The practical evaluation of \flex shows that it enables the application of SMC in dynamic environments with reasonable performance penalties and computation durations allowing soft real-time and interactive use cases.

\end{abstract}
\IEEEpeerreviewmaketitle

\section{Introduction}
The Internet of Things (IoT) and Smart Buildings are emerging paradigms which aim for joining the physical and the digital world.
An essential step towards this is to deploy a multitude of sensors in physical environments which then allow gaining insights into the environment's current state \cite{Gubbi2013}.
This data can be used to provide a fine-grained understanding of real-world processes, inform users,  and (automatically) influence and manage the physical environment\onlylong{ in the sense of a feedback loop. 
  It is immediately obvious that the availability of this data is essential for the IoT and Smart Buildings to function}.
State-of-the-art cloud-based and local but centralized data stores, however, bear similar data protection problems:
They constitute a single point of attack for all users' data.
Likewise, conformance with data protection regulations  
can be a complex and challenging task. They might even completely prohibit valuable data usage, createing a trade-off between security and utility of collectable data.

SMC is a technical solution to these conflicts.
Its application is possible due to the following observation:
Often, collected data is only needed in post-processed form.
Data points of different times or different locations are combined and
 numerical values are transformed into categorical (or boolean) decisions\onlylong{ by operations like thresholding}.
While the raw input data might have been privacy-critical, often, the final output is not.
The technology has already proved applicable \cite{SEPIANetwork}\!\cite{SMCFinancial}\!\cite{SugarBeetSMC, SMCOutage, SMCAviation, Bonawitz2017} (cf. Section~\ref{sec:related_work}).
However, publications stress that more focus should be laid on the practical problems of applying SMC:
Administrative real-world challenges should gain more attention and more challenging settings like cloud-environments should be considered.
Bogdanov \etal \cite{SMCFinancial} ``consider it an important challenge to reduce the administrative attention required for managing'' a computing node to make ``the technology easier to deploy in practice''.
Furthermore, Burkhart \etal \cite{SEPIANetwork} recognize that robustness of computations in the presence of host failures is a vital challenge which should be considered in future.
We address exactly these problems in our work.

We capture this demand by focusing on smart environments as use case.
They are understood to be dynamic settings where a completely dependable infastructure cannot be assumed.
Similarly, host and networking failures have to be taken into consideration when designing software for them.
Nevertheless, the need for privacy-preserving computation is strongly given, which is preferably performed repeatedly and automatically without notable overhead of manual management.
We use smart office buildings as running example:
Information about building usage is distributedly collected in every office space. It is used to provide insights for building managers as well as the inhabitants (e.g. via public displays).
Furthermore the same data can be forwarded to controllers and actuators influencing the state of the building (e.g. HVAC).
However, privacy-preserving processing is needed as the collected data especially provides data about the employees working in the sensed area.
It can contain presence information about individual employees and give insights about their working and moving behavior. Tracking and profiling becomes possible. \cite{Roman2013}

In the following, we explicitly address the problems which emerge when trying to apply SMC in these contexts.
Our contribution is
the wrapper framework \emph{\flex} for SMC implementations
which enables managing of participating nodes and the
orchestration of collaborative computation.
It facilitates setup including node discovery and cryptographic bootstrapping\onlylong{ making authenticated secure point-to-point channels available}.
During computation it performs monitoring and alive checks of the peers and provides mechanisms for session recovery in case of failures.
As a result, \flex enhances SMC to be used as a dependable und self-managing service services in dynamic and unstable environments.

We structure our work as follows: In
Section~\ref{sec:background} we provide background about SMC.
Section~\ref{sec:challenges} outlines the challenges for SMC in dynamic contexts while
\ref{sec:virtual_centrality} discusses how they can be conceptually approached.
In Section~\ref{sec:overview} we give an overview over the architecture which is then described in detail in
\ref{sec:arch}.
Section~\ref{sec:eval} presents our evaluation showing the results of our performance measurements.
Section~\ref{sec:conclusion} concludes the paper.

\section{Secure Multiparty Computation}
\label{sec:background}
Secure multiparty computation (SMC) is a method which allows a group of parties $P_1 \ldots P_n$  to collaboratively compute a common function $f$.
Hereby, each party $P_i$ can contribute an input $x_i$ to the computation; no other party learns this value but all learn the result $f(x_i, \ldots, x_n)$.

Yao \cite{Yao1982}\!\!\cite{Yao1986} was the first to question how two parties $P_i$ and $P_j$ can privately compare their values $x_i$ and $x_j$ so that both only learn which one is bigger without sharing the actual values.
Later, this question was generalized to arbitrary computations.
This resulted in the research field of SMC.
Since then, research focused especially on feasibility of SMC in a variety of contexts, increasing its performance and improving its security in stronger adversary models.

SMC is executed as protocols between the participants.
These protocols are typically organized in synchronized, sequential rounds.
A round consists of a computation step, where every node performs the same calculations predefined by the protocol and
often a communication step where some data is exchanged to proceed with the next round.
This communication has to happen between every pair of participants.

\section{Challenges for SMC in Dynamic Environments}
\label{sec:challenges}
The Internet of Things and smart environments bear characteristics which are initially incompatible with SMC.\onlylong{ These are the challenges which have to be solved in order to apply SMC successfully.}
We consider solving these conflicts as the fundamental requirement in order to transform SMC into a valuable service in smart environments.
The main conflicts are:

Before executing an SMC protocol, the nodes need to know each other \req{req:know}.
They must have an established trust relationship in order to create secure authenticated point-to-point channels \req{req:secure_channels}.
Compatibility for computation \req{req:compatibility} between nodes must be ensured: They must be capable of the same SMC protocols for being syntactically able to interact and provide the same type of data for producing semantically sensible results.
Furthermore, during the computation, the set of participants cannot dynamically change \req{req:static_set}.
Coordination of the SMC protocol has to be performed from outside \req{req:trigger}, at least the initiation of the computation has to be triggered at each node nearly synchronously.
SMC is not necessarily robust against failing nodes during computation and recovery of sessions is not trivial. Consequently, the environment where SMC is executed must be robust itself \req{req:robust_env}.
Lastly, SMC technically uses peer-to-peer protocols, they are initially incompatible with a client-server architecture \req{req:client_server}. I.e. clients cannot trivially request data from such a network. Furthermore, at the end of an SMC computation the result is available for each participant and no one outside this group. If other entities need access to the result it must be forwarded \req{req:result_forwarding}.

Smart environments on the other side are dynamic, i.e. nodes are not initially known to each other, no trust relationship between them exists initially,  and nodes can unexpectedly join and leave the network\onlylong{ (deliberately or due to failure)}.
Every node might have different sensors attached, computation compatibility is not trivially given.
Lastly, in an environment where other services shall depend on the results, a single point of contact is desirable.

\section{Approach: Virtual Centrality}
\label{sec:virtual_centrality}
When formally proving security and correctness of SMC protocols, a \emph{simulator-based} proof approach \cite{Canetti1999} is used. Roughly spoken, the proof shows that an adversary cannot differentiate whether it interacts with an  SMC protocol or an actual Trusted Third Party (TTP).\onlylong{ That means, SMC protocols are secure and correct if they behave like and leak only as much information as TTPs.}

We use this model as inspiration: TTPs bear characteristics which are desirable to address the management challenges mentioned in Section~\ref{sec:challenges}.
Hence we try to incorporate these properties while using SMC as core.
Given by the scenario, there are distributed nodes acting as data sources. In the following, we denote them \emph{peers}.
Similarly, data consumers exist, they are \emph{clients} of available services, like public displays or HVAC controller as exemplified before.
For our solution, we now define a special further node -- the gateway (GW) -- which acts as a virtual server,  shadowing the fact that data is actually stored in a distribued manner and requests are answered by collaborative on-the-fly computation by the data sources. The GW performs centralized coordination and orchestration of SMC computations but is not trusted with respect to the data. We call this concept \emph{Virtual Centrality}.

\image{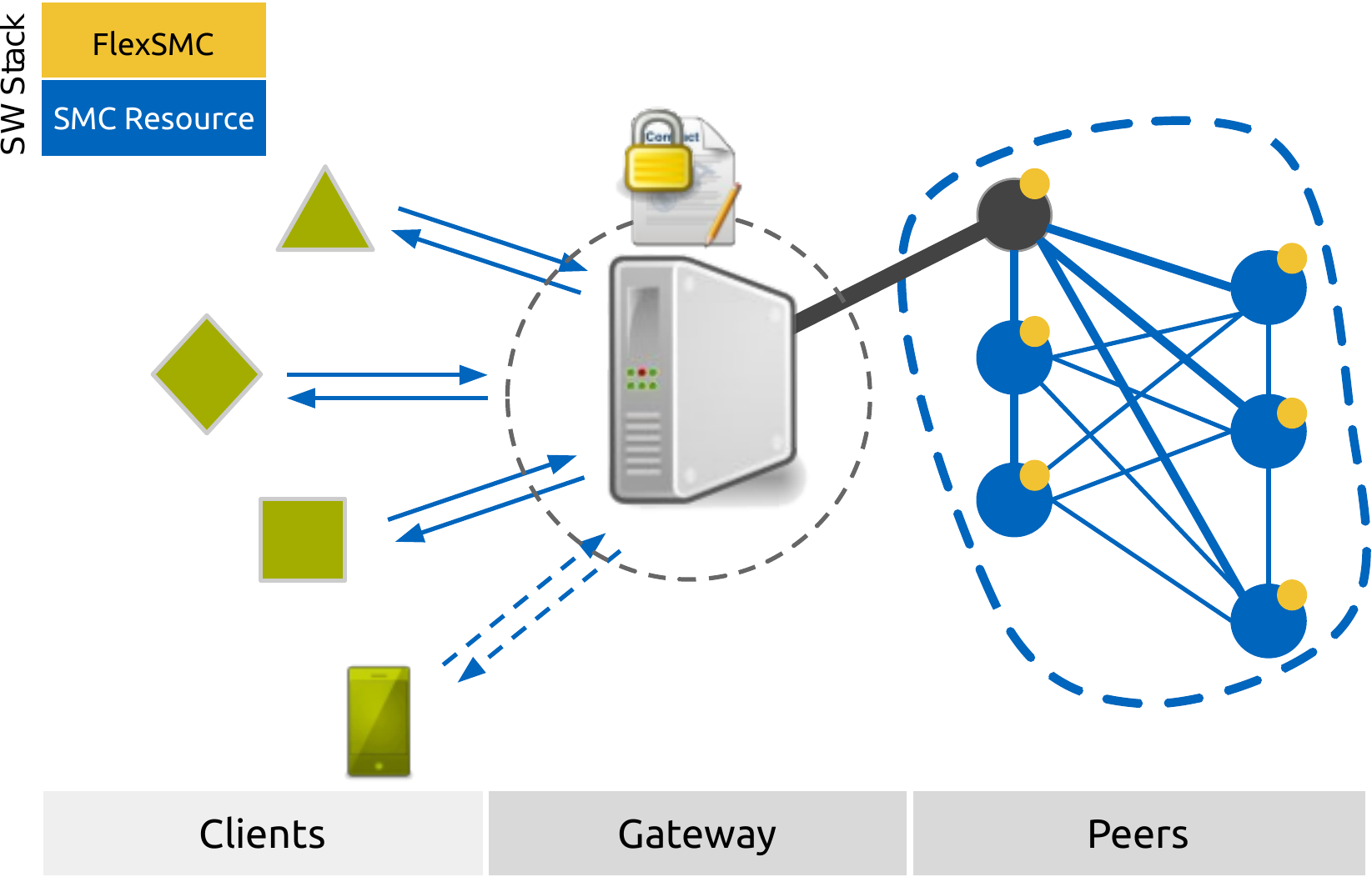}{Overall architecture}{fig:overall_architecture}
The main functionality of the GW can be divided into two parts (cf. Figure~\ref{fig:overall_architecture}):
Firstly, it provides a unified API for data consumers, hiding SMC from them.
Clients can request preliminary metadata which nodes and which data are available.
Based on this information, clients can issue computation requests, i.e. requests of data which is then newly computed by a corresponding SMC session between the peers. Clients finally receive their answers via the API.
Secondly, it must coordinate the computations performed by the data sources.
It acts as a discoverable management node which initially collects metadata about the capabilities of the connecting peers and establishes control connections to them.
Upon a computation request, it initiates and orchestrates a corresponding computation, finally receiving the result itself but nothing else.
Due to the mutual trust relationship among all nodes, and discovery and bootstrapping mechanism, any node is potentially eligible for the gateway's leader role.
In use cases where a preliminary infrastructure can be assumed (e.g. smart offices or homes), the gateway can be assigned rather statically to a dedicated node. In highly volatile environments like ad-hoc-networks, the gateway role can be chosen completely dynamically at runtime.

\section{Overview}
\label{sec:overview}
In the following section we present the overall architecture of \flex, our approach to practically realize Virtual Centrality and address the challenges discussed in Section~\ref{sec:challenges}.

\subsection{Entities}

We elaborate further on the roles of entities implied by our use case and added by our solution:
\subsubsection{Peers}
\emph{Peers} are sensor platforms which gather raw information.
They can differ regarding their capabilities, i.e. the attached sensors.
This and other information are available at the peers as metadata.
The peers carry out the SMC sessions.
\subsubsection{Gateway}
The \emph{gateway} is a central yet untrusted \emph{peer}-like node. It functions as described in Section~\ref{sec:virtual_centrality}. It presents itself as a monolithic service for all clients and coordinates and orchestrates the SMC sessions for all participating peers.
\subsubsection{Clients}
\emph{Clients} are data consumers which request and finally receive the collected data to perform actions upon them or make them visible to users.

\subsection{Functionality}
We implement a GW which provides functionality to cover all problems stated before in  Section~\ref{sec:challenges}.
It contains an orchestration module which solves \ref{req:know}, \ref{req:secure_channels}, and \ref{req:compatibility} during a setup phase for every joining peer.
The monitoring module addresses \ref{req:static_set}, \ref{req:trigger}, and \ref{req:robust_env}.
The GW will also participate in the computations and also obtain the final result. I.e, after having received a request and translated it into an SMC session (\ref{req:client_server}), it is also able to response with the result at the end (\ref{req:result_forwarding}).

\flex acts as a wrapper for existing SMC frameworks. I.e. we do not duplicate SMC functionality but provide an environment which mitigates the conflicts between dynamic contexts and the premises of SMC application.
\flex then connects via a local socket to an SMC instance and controls it via an adapter  (cf. Figure~\ref{fig:eval_metric}). This decouples \flex from specific SMC implementations.
This is also beneficial with respect to improvements on SMC protocols. As long as no fundamental changes in the structure of sessions are made, \flex remains compatible.


\section{Architecture}
\label{sec:arch}
In this section, we explain the aforementioned modules of \flex in detail.
\subsection{Orchestration}
When new nodes join they need to find an appropriate GW instance.
This GW then has to gather information about the new node,\onlylong{ get to know its capabilities,} and prepare it for later computation sessions. The peer-side process is depicted in Figure~\ref{fig:peer_state}.

For \emph{discovery} of
GWs we employ mDNS \cite{rfc6762}\!\!\cite{rfc6763}.
The GW acts as a service which announces itself over the network.
Since multiple GWs can be present which may have different locations, purpose,   etc., GWs also send this metadata during this announcement.
This facilitates selection of the right GW for new peers.
Selection can happen manually or automatically on the peer-side.
When a peer has selected a well-suited GW, it establishes a connection to the GW and starts the pairing.

During \emph{pairing},
self-signed certificates of the new peer and the GW are exchanged which
provide cryptographically secure identities for the
peers.\footnote{
This can be substituted by a Public Key Infrastructure if available.
}
\image{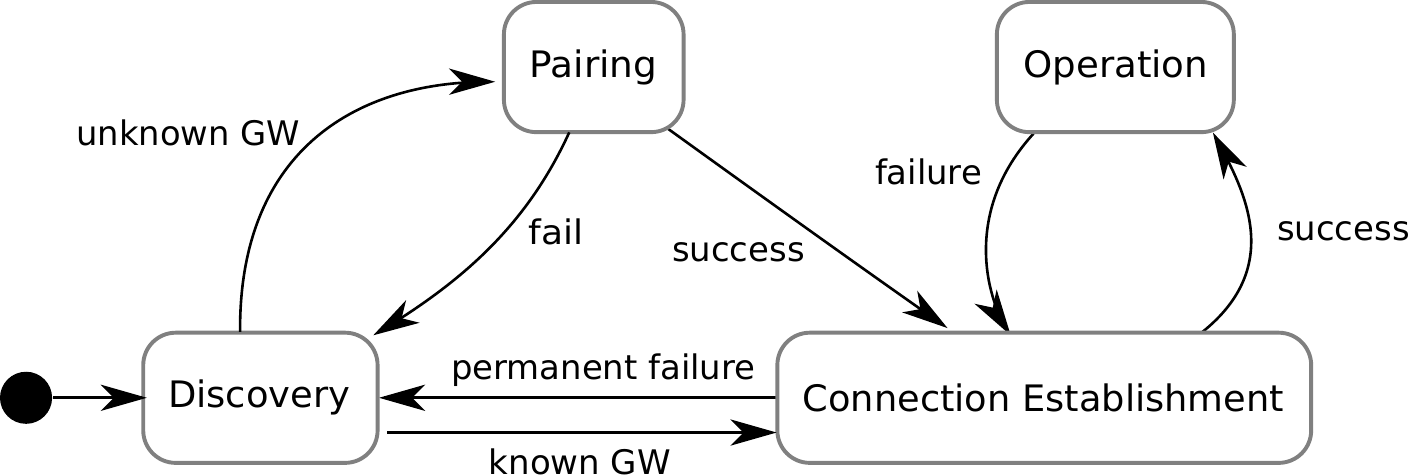}{State machine of peers}{fig:peer_state}
Later, when invoking a computation, all necessary certificates are distributed among the participating peers in order to enable computations performed over secure channels.
Furthermore, the peer provides metadata which enables creation of semantically sensible SMC groups of peers. Examples for this data are the peer's location and its capabilities.
Each group gets a label which enables clients to address this very group by its name.
It is the task of the GW to resolve the name again when a computation is requested.
Lastly, when \emph{connecting} a permanent control channel is established to enable the GW to later provide instructions to this peer.

When \emph{operation} has been started,
peers are ready to execute computations.
Incoming requests from clients are preprocessed by the GW.
The group label is resolved and the input data type and the choice of protocol are extracted. Afterwards, the GW informs the corresponding set of peers about the upcoming computation and its metadata. This includes the identities and connection endpoints\onlylong{ (IP address, port)} of all other participants. Peers can then prefetch the input data and perform local setup of the SMC implementation.
On each peer, the computation is delegated to a dedicated SMC implementation. It uses the metadata to connect to the participants and collaboratively execute the selected protocol.
When the computation has finished, all participants, including the GW, have obtained the result. The GW forwards the result to the initially requesting client.

\subsection{Monitoring}
Monitoring mainly observes the state of and the connection to already joined nodes.
\subsubsection{State Monitoring}
\label{sec:state_monitoring}
The availability of a joined node is monitored in a two way fashion:
Regularly, peers send heart beat messages to the GW.
When the connection drops, peers will take notice of it, clean up the connection and transition back into discovery mode (cf. Figure~\ref{fig:peer_state}) trying to reestablish a connection the same or -- in case of a permanent error -- to another GW.
On the side of the GW, missing heart beat messages or connection loss in the control channel indicate errors which also cause the cleanup of the corresponding channel.
The peer is then unlisted as active and available.

\subsubsection{Computation Monitoring}
We delegate the computation to a SMC framework which we assume to be available on each peer.
As a consequence the actual computation is carried out via other channels than the channels \flex controls.
Therefore, SMC communication cannot be monitored directly.

However, \flex is connected to the SMC framework so that exceptions can be retrieved from the computation session.\onlylong{ This is possible as a peer or connection failure is visible for every participant; i.e. also the GW.}

Assuming that SMC sessions cannot recover themselves, the GW then instructs to cleanup the failed session and to begin recovery.
Using the information from state monitoring (Section~\ref{sec:state_monitoring}) the GW optionally removes vanished or failed peers and restarts the computation a predefined, finite time.

\section{Performance Evaluation}
\label{sec:eval}
\image{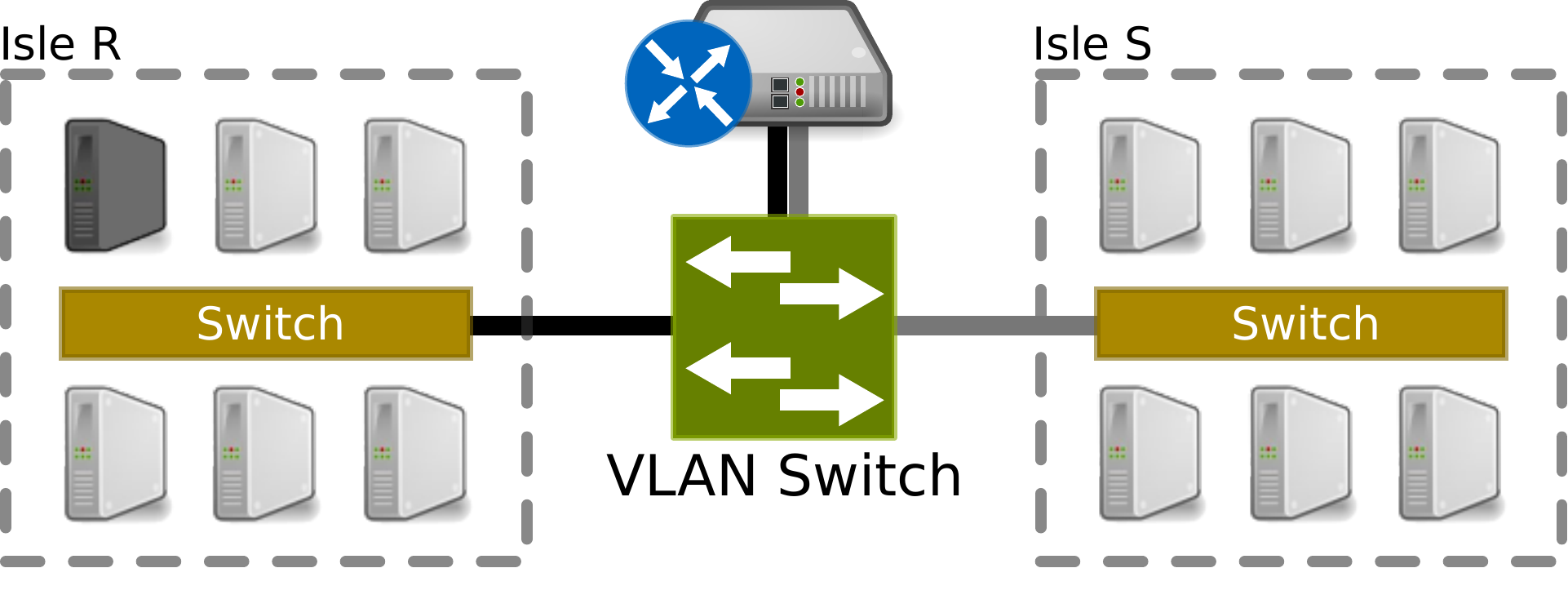}{ Topology of the test setup }{fig:evaluation_setup}
In order to assess feasibility of our solution we performed measurements focusing
on absolute durations for the processing of requests and the overhead induced by \flex.

\subsubsection{Setup}
We used 12 physical hosts in a topology as depicted in Figure~\ref{fig:evaluation_setup}.
There is a switch per subnet; both are in turn connected by a central switch.
Traffic between subnets needs to pass the router.
The left upper node of Subnet R acts as a dedicated GW\onlylong{ and orchestrates the other 11 nodes}.
Each host possesses 16\,GB main memory and an Intel Xeon CPU with 8 cores having 2.50\,GHz and 8192\,KB cache size.
The operating system is a headless Debian 8.5 Jessie using a Kernel of version 3.16.0-4. They are connected via a 1\,Gbit ethernet network.

We employed \FRESCO \cite{FrescoURL} (version 0.2) as SMC framework. The basic computation operations are provided by the BGW protocol \cite{BGW88}.
Being guided by the smart office example, we perform a simple summation of all participant's input values as it is typically necessary for generating statistics from collected information, e.g. calculating the average.
Each measurement has been repeated 1000 times.

\subsubsection{Method}
\image{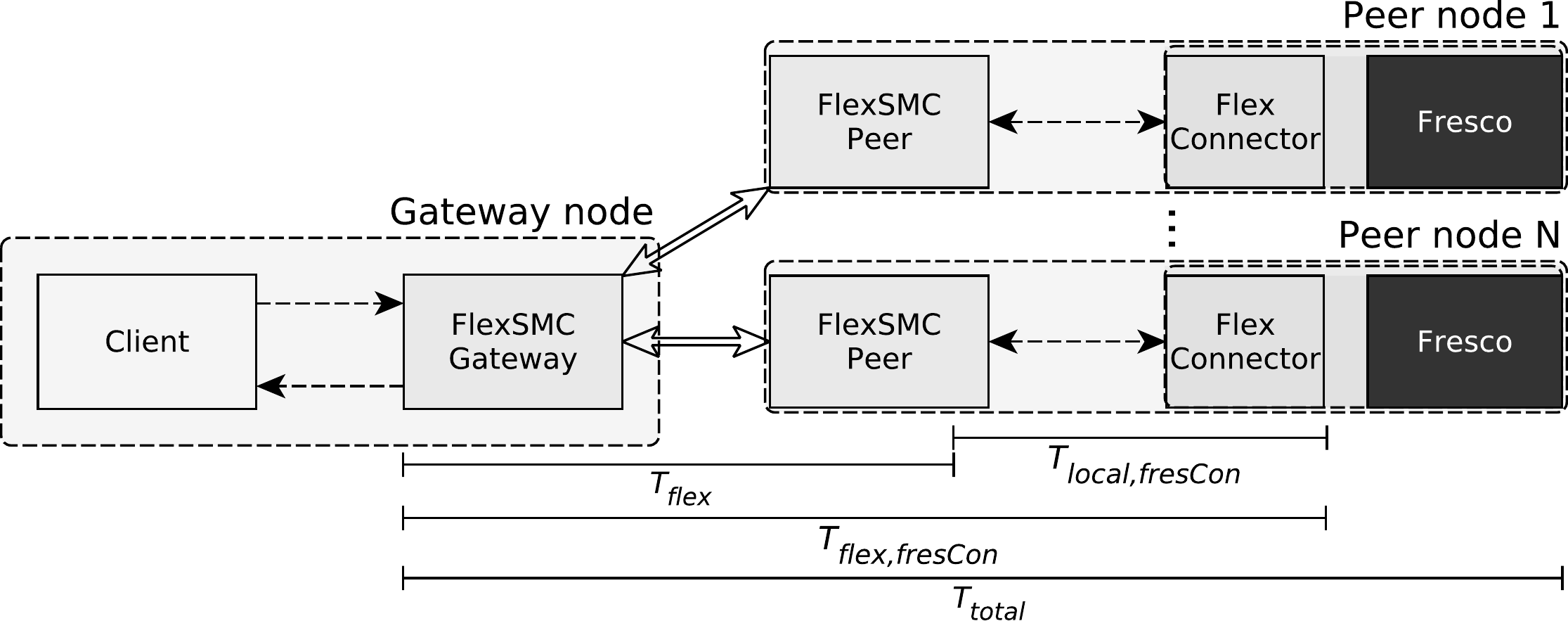}{Test metric for the complete communication path
}{fig:eval_metric}
We dissect the steps from request to computation corresponding to Figure~\ref{fig:eval_metric}, gaining different test cases.
Here, $T$ always denotes the time for a bi-directional communication flow.
Dotted edges visualize local communication on a host while solid edges between \flex instances  depict any routed and switched network.
\(T_{flex}\) denotes the time which is needed to inform peers about an upcoming computation \onlylong{by providing the necessary metadata }and to invoke the computation on the peers.
Adding the time spent while communicating via a local socket with the SMC implementation yields \(T_{flex,fresCon}\).
The full time taken, also including the actual execution of the SMC protocol, is \(T_{total}\). In this work, we will provide measurements for \(T_{flex}\) and \(T_{total}\).

Communication is carried out simultaneously with every peer. However, during the SMC computation each round has to be synchronized among all peers. Practically, this means that the slowest peer determines the overall performance.
Due to this reason all our results always depict the maximum duration, i.e. the time the slowest peer took per measurement.

\subsubsection{Results}
We provide three insights:
a) The overall time a protocol execution costs. b) The fraction \flex adds to this amount. c) The scaling behavior with respect to the number of peers, an important parameter in our setting.

Figure~\ref{fig:phases} shows that a sequence of 10 consecutive ``echo'' requests sent to the \flex
peers via individual control channels
involving all \flex framework components in an end-to-end view
but without connection to the local adapter and actual computation costs around 4\,ms.
This value increases very  slightly when more peers are added.
Actual forwarding of these messages via the local socket adds further 3\,ms to the overall duration.
Regarding scaling behavior, the offset is correspondingly shifted and the slope is increased.

\image{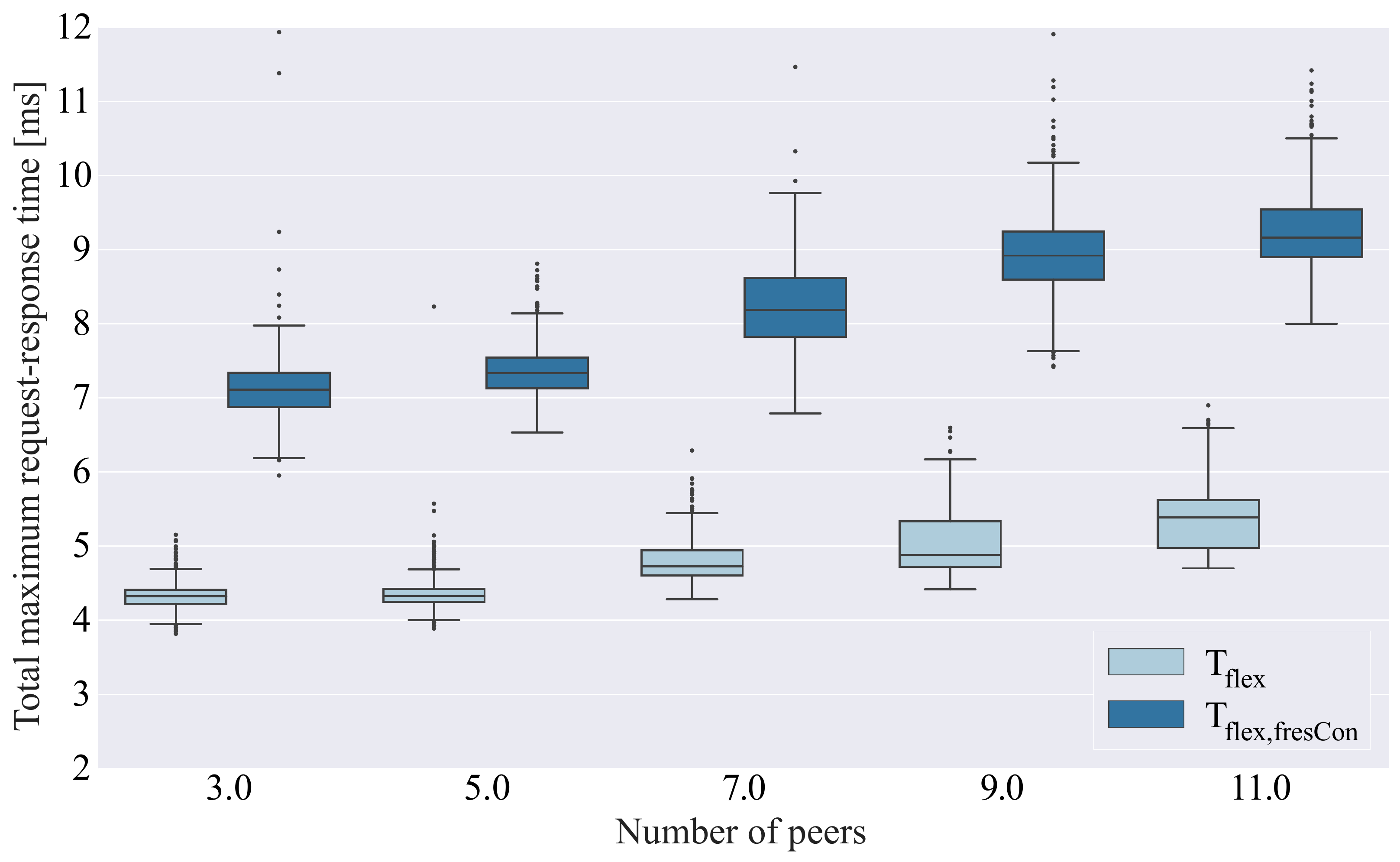}{Total time for 10 consecutive echo requests}{fig:phases}
\image{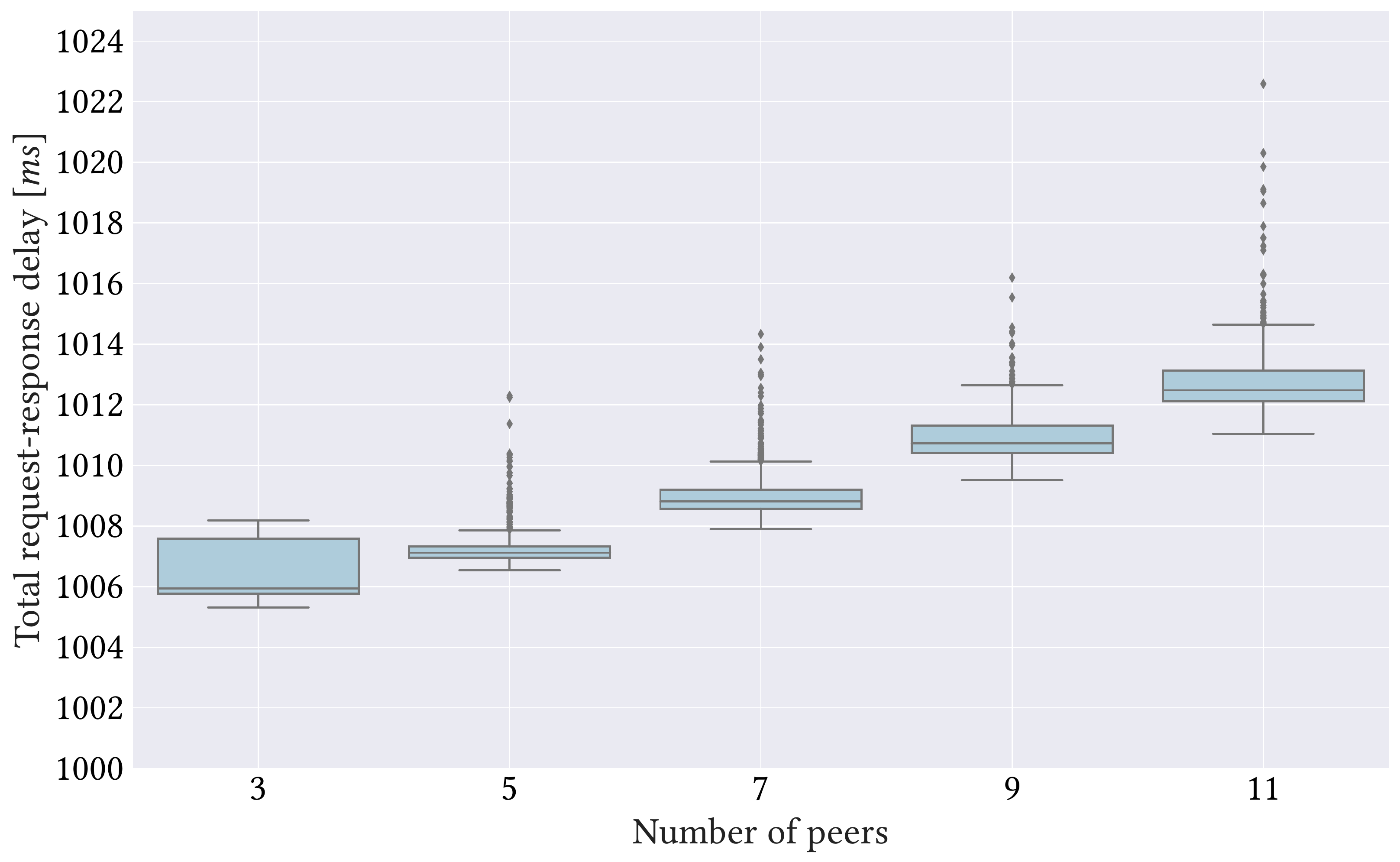}{Computation of a single secure sum per round with different number of participants\onlylong{. The result represents an end-to-end perspective on the system measuring the elapsed time for starting a request, performing complete necessary peer interaction, its network round-trips, the preparation and execution on \FRESCO side, and returning the result back to the client.}}{fig:e2e_msmt}

The next measurements included the actual SMC computation.
Figure~\ref{fig:e2e_msmt} shows that the corresponding durations have an essentially higher offset. The setting with 3 peers starts with 1006\,ms and increases to 1012\,ms for 11 peers.
Investigating this issue, we found out that this is neither SMC-specific nor directly caused by the performed computation. Instead, the networking layer of \FRESCO always waits a full second before checking whether all necessary channels between each pair of peers has been established.
This is also the reason why this is only an offset which does not influence the slope.

Besides absolute durations comparison with a TTP solution as given in the state-of-the-art is of interest.
We can approximate its cost by assuming that the GW would act as a TTP, performing the computations centrally.
For this purpose, each peer would forward its data to the GW.
Using this approximation and neglecting the central computational costs, utilization of a TTP would then be around \(T_{flex}\) since its corresponding commmunication is not SMC-specific.
It is, however, vital to see that the costs would be nearly independent of the number of peers contribution to the computation.
All peers would be able to send their values simultaneously.
This is in stark contrast to the SMC solution:
Carrying out the SMC protocol requires communication between all participating peers. This causes a linear increase of the duration when adding further peers.\footnote{The initially quadratic overhead is reduced to linear due to parallel communication.}

Due to the waiting time in \FRESCO, client requests would be answered in roughly more than a second.
Here, \flex imposes only minimal overhead, and with further versions of \FRESCO,
the overall time can even decrease to some milliseconds.
These performance characteristics enable serveral use cases:
Batch data processing and interactive applications become feasible.
Similarly, continuous applications requiring soft real-time like public displays or actuators for HVAC control would be well realizable.

\section{Related Work}
\label{sec:related_work}
Over the last decade there have been multiple feasibility studies demonstrating practical SMC application.
However, it becomes apparent that most solutions presented in related work have been executed manually in highly controlled environments. Only few propose architectures for continuous real-world deployment and automated computations. Among them, some in turn do not consider realistic constraints of their environment.

In constrast, we deliberately chose dynamic environments and automated computation as our scenario in order to examine and address infrastructural challenges for SMC present in smart environments. We provide a solution which allows continuous and automatic application of robust SMC computations in dynamic contexts.
We evaluated our prototype in a real network and proved its feasibility for practical purposes.

The first practical and large-scale application of SMC happened in 2008 \cite{SugarBeetSMC}.
Its purpose was to perform a multiple seller multiple bidder auction\onlylong{ for computing a market clearing price\footnote{Given multiple prices where sellers are willing to sell a certain amount and where bidders are willing to buy a certain amount of a commodity,  then the market clearing price is the price where the total demand equals the total supply. }}.
\branch{The data was collected from 1200 users\onlylong{ without special technical experience by} using a Java applet in a web browser.
It was separated in shares locally and each share was encrypted for one of the three computing parties.
They were then sent to a single central collection server.
At the time of computation, three laptops were used as computing parties being connected in a local network. Every party owner obtained his shares from the collection server, decrypted his shares locally by entering the password and then started the computation manually.}{Data from 1200 users was collected and split into shares locally in a Java applet. The computation has then been performed by three laptops in a local network.}
The whole computation, working with 1229 bids encompassing 9 million individuals numbers took 30 minutes in an 100Mbps intranet setting.
They do not give insights how the results were distributed to the initial data input parties.

%
%
Martin Burkhart\branch{ applied SEPIA -- developed by him during his PhD \cite{Burkhart2011} -- }{ developed and applied SEPIA} to event correlation for network data \cite{SEPIANetwork} in 2010.
The overall goal is generation of network traffic statistics and anomaly detection.\onlylong{This was realized by computing histograms, entropies and performing distinct counts of input values.}
The collected data stems from 140 input parties while varying the computation parties between 3 and 9. \onlylong{Working with 65000 inputs per input node in a 100 Mbps intranet setting, all statistical computations took around 1 to 2 minutes. }Their evaluation was performed in a very controlled setting: Measurements have been performed in intranet and internet settings (using PlanetLab), however without real interaction via SMC with the ISPs which provided the data.

%
Bogdanov \etal \cite{SMCFinancial} set up SMC in 2012 for computing statistical financial indicators of\onlylong{ information and communication technology} companies.
\onlylong{These give helpful insights and the ability for self-assessment but are based on confidential company information. }Sharemind \cite{SharemindPhD} was used for realizing 
\onlylong{an Oblivious Batcher's odd-even merge sorting network besides other functionality for performing }ranking operations.\onlylong{ Their protocols are hence based on additive secret sharing, the number of computation parties is 3.
 The number of input parties is not documented, but a questionnaire performed along their work implies that there are around 15 -- 30 data input parties.}
They provided a web-based solution which created the shares locally in the browser via JavaScript before submitting them to three computational nodes.
These were located in three participating companies\onlylong{ having the necessary knowledge to maintain such an instance}.
They successfully deployed it as an application for continuous use, several computations have been carried out.
%
%

Djatmiko \etal \cite{SMCOutage} reused SEPIA in 2013 to perform collaborative outage detection.
\branch{
They used the existing \emph{Flow-based Approach for Connectivity Tracking} -- which typically works with a single input source without privacy considerations -- 
and extended it}{They extended an existing approach} to work privately with multiple inputs.
The core operation is multiset union operation which they realize with \emph{counting bloom filters (CBFs)}.
Their CBF works on an integer array\onlylong{ of length 32.768,} which has to be generated from equally-shaped, local arrays.
The necessary summations of the interger elements are performed via SMC.
Real-world data from an ISP is used and evaluated in a controlled and manually setup cloud setting.

In 2016, Zanin \etal \cite{SMCAviation}  applied SMC to\onlylong{ an auction method implementing the EU Emission Trading Scheme.
Herewith, they made it possible to} transmit \(\text{CO}_2\) emission allowances between airlines.
They used SEPIA implementing \branch{a comparison protocol by Nishide and Ohta \cite{Nishide2007} to yield the necessary auction scheme}{the necessary comparison functions}.
Most notably, they have realized a real web service interacting with actual stakeholders.
\onlylong{Representatives of airlines can enter their bid and an external referee, acting as auction manager operates their auction system. }While the interaction is neither fully automated nor the SMC computations are performed automatically, they provide an infrastructure which addresses a concrete use case and allows real interactions.
The same is reflected in their measurements, where \onlylong{not only the SMC computations but }also organizational overhead is assessed which includes peer discovery and authentication.

In 2017, Bonawitz \etal \cite{Bonawitz2017} from Google published a solution for SMC-based privacy-preserving training of neural networks (NN).
\branch{The trained model of a NN consists of the adjusted weights of the nodes of the network which}{A trained model of a NN} can be represented as a (long) integer vector.
Training such a model \emph{privately} means that a global vector is build from a multitude of local input vectors without making these available to anyone.
In their approach they address real-world problems like interrupted connections, vanishing participants, and NAT-shielded devices.\onlylong{ They hence require that their protocol works asynchronously, a single faulty participant cannot jeopardize the full computation and that a centralized architecture is sufficient which only requires communication of each input party with a single server.}
The corresponding talk \cite{Kreuter2017} implies that the system will be productively used
for privately creating auto-suggestion models for smartphone input trained by the typing history of a multitude of smartphones.
Abstractly, Bonawitz \etal address \emph{Secure Aggregation}.
This facilitates their use case:
\branch{
They do not need several computing parties which are able to carry out arbitrary computations. In contrast, they
}{
They
}
only need a single server finally holding the global model while being oblivious to the input data. Furthermore, only integer summation must be supported.

Thoma \etal \cite{Thoma2012} present a secure smart metering solution.
They aim for protecting the individual, temporally fine-grained consumption data of households, providing correct and verifiable billing of each individual household and fine-grained (non-individual) consumption feedback for load management.
\branch{They achieve this by a dual approach. Temporally fine-grained feedback data is collected by aggregating individual consumption data before sending it to the energy provider. This is performed via SMC and individual consumption data is protected. Monthly aggregates for billing are created in plain and locally for each household. This values become verifiable as every household sends individual, fine-grained but homomorphically encrypted consumption data to an untrusted  \emph{utility server}.
The server can sum up the inputs and use it for secure verification of the user-side result provided at the end of the month.
}{Here, SMC is utilized to spatially aggregate temporally fine-grained consumption data over several households before sending them to the energy provider.}
Thereby, they also address a secure aggregation in a real-world setting and provide a framework which enables practical application.
However, their solution is rather on the level of a concept.
The paper does not discuss how infrastructural problems like discovery of available households and interconnection between them are handled.

\section{Conclusion}

\label{sec:conclusion}
Smart environments and the Internet of Things are emerging technologies which require the privacy-protection of distributedly collected data.
Secure multiparty computation (SMC) is a valuable candidate for this purpose.
Its
practical feasibility has already been shown in several studies in the last
decade. However, most of them have been carried out in highly controlled
environments and with a lot of manual intervention. Relevant questions
of productive application in dynamic environments have therefore not yet been answered.

We examine the problems arising for SMC when considering these contexts:
Due to their volatility preliminary information about possible participants and their identities are not necessarily known. Similarly, secure channels for SMC require previously performed trust exchanges between the participants. Lastly, failing hosts and connections must be taken into consideration, with which SMC realizations cannot trivally cope themselves.
Our contribution is \flex, a management and orchestration framework which turns SMC into a dependable service for these scenarios.
Its main paradigm is \emph{Virtual Centrality}: A selected node among the data sources provides management and orchestration functions for enabling computations while remaining untrusted with regard to data handling.
During setup, it supports node discovery and trust establishment.
It serves as single point of contact for data requests and translates them into executable SMC sessions. When performing computations, it monitors them and enables recovery when node or communication failures occurred.

Our performance evaluation shows that SMC in general is feasible in dynamic environments and \flex adds only negligible overhead to it.

\bibliographystyle{IEEEtran}
\bibliography{literature,rfc}

\end{document}